# New Thermodynamic Paradigm of Chemical Equilibria

B. Zilbergleyt [1]

> "Nature creates not *genera* and *species,* but *individua,* and our shortsightedness has to seek out similarities so as to be able to retain in mind simultaneously many things."
> *G. C. Lichtenberg. "The Waste Books", Notebook A, 1765/1770.*

## 1. Introduction.

These words, though ironically, express the essence of the Ockham's razor [1] − to employ as less entities as possible to solve maximum of problems. Both reflect the efforts of this work to unite on the same ground some things in chemical thermodynamics that seem at a glance very unlikely.

A difference between theoretically comfortable and easy to use isolated chemical systems and real non-isolated systems was recognized long ago. A notion of the open chemical systems non-ideality, offered by G. Lewis in 1907 as a non-ideality of species [2], was a response to that concern. Fugacities and thermodynamic activities, explicitly accounting for this notion, were introduced for non-ideal gases and liquid solutions to replace mole fractions in thermodynamic formulas. Leading to the same habitual linear dependence of thermodynamic potential vs. activity logarithm (instead of concentration), that allowed the thermodynamic functions to keep the same appearance, passing open and closed systems for isolated entities. With this support classical theory has survived unchanged for a century. J. Gibbs is credited for a presentation of phases in the multiphase equilibrium as open subsystems with equal chemical potentials of common components [3]. His insight hasn't impacted essentially thermodynamic tools for chemical equilibria.

First stated in a general way by Le Chatelier's principle, energy/matter exchange is forcing the open system to change its state. Being an adaptation of Gauss' least constraints principle, Le Chatelier's principle has linked thermodynamic stresses, imposed on the chemical system by external impact, with the system response: to decrease thermodynamic mismatch with its surroundings, stressed open system flees to a state with minimal stress, allowed by external constraints. The principle describes perhaps the simplest known self-organization in nature. H. Le Chatelier (see [4]), R. Etienne [5], T. De Donder [6] tried to move that verbal principle towards the numerical form, offering a set of moderation theorems. Unfortunately, their efforts ended up only with inequalities. Self-organization and dissipative structures in open chemical systems are extreme manifestations of the principle, but their quantitative description is well beyond the grasp of classical thermodynamics. The term "far-from-equilibrium" was coined to designate loosely a place of honorable exile from equilibrium thermodynamics for a great deal of real systems. Chemical thermodynamics, seemingly powerful and definitely elegant in its traditional applications, became torn apart and turned into a clumsy schizothermodynamics ("split thermodynamics"). It is literally trying to apply different concepts to traditional isolated systems with true thermodynamic equilibrium as the only achievable state and to open systems with complex behavior − none of the recognized models allows for transition from one to another without sacrificing model's integrity. At this point classical chemical thermodynamics looses its power. In classical theory, isolated system is a heavily reigning predominant routine, sometimes releasing the vapor through the idea of closed systems, while open systems actually never arrive. Isolated entities are treated within the classical "energetic" concept; treatment of the self-organizing systems employs mostly the "entropic" approach. Works of I. Prigogine and his school regarded the entropy production as the major factor, controlling the chemical processes, actually substituting energy and putting entropy in charge of the system behavior. Some authors consider the "entropic" concept more fundamental than the "energetic" [7]. According to [8], the energetic treatment makes more sense in isothermic systems, otherwise the entropic one is more preferable.

---

[1] System Dynamics Research Foundation, Chicago, USA, e-mail: sdrf@ameritech.net.

Also, the latter is not good for systems with robust reactions, and cannot cover the full range of chemical transformations. Discrete Thermodynamics of Chemical Equilibria (DTd), the subject of this paper, based on a new paradigm of chemical equilibria, represents a unique theory on the energetic basis.

## 2. Definitions.

### 2.1. Explanatory notes.

Like mathematics, chemical thermodynamics doesn't care about names; the only values that matter are energy, entropy and corresponding thermodynamic parameters, and amounts of species. Taking advantage of that, we based our derivations and the most of examples mainly on the abstract symbolic stoichiometric reaction equations like A+B=AB. Unless specified, the simulations were carried mostly at p=100kPa, T=323.15 K, these values were taken just for being the same in all examples without any specific reason; a set of the $\Delta G^0$ values for the real gas reaction

(1) $\qquad\qquad\qquad\qquad\qquad\qquad\qquad\qquad\qquad\qquad\qquad\qquad PCl_3+Cl_2=PCl_5$

was taken as basic just to have realistic numbers. In reality the set covers $\Delta G^0$ range −31.18 to 27.04 kJ·m$^{-1}$ and T from 323.15 to 673.15 K, we used various (mostly negative) $\Delta G^0$ at the same 323.15 K.

The goal of chemical thermodynamics is to predict behavior of chemical systems in terms of their thermodynamic states. Being expressed in terms of the system shift from "true" or internal thermodynamic equilibrium (TdE), their response to the external thermodynamic forces plays the key role in DTd. The shift takes on either positive or negative values; both can be modeled directly. At the same time, negative shift of direct reaction corresponds to the positive shift of its reverse reaction, and such an approach to the 2-way modeling occurred to be more smooth and rational. Later on the reverse shifts are represented conditionally by negative values in the graphs. The simulation methods and a series of computer programs were developed to find numerical solutions and have been permanently improved as soon as new ideas were tested and survived during last decade; the basics of the programs is briefly described in Appendix A. Unless else stated explicitly and besides the chapters, relevant to the electrochemical systems, through the whole paper the graphs are plotted in $\delta_j$ (shift, ordinate) vs. $\tau_j$ (growing factor, abscissa) coordinates. Parameters, shown on the upper white field of the pictures and on the captures follow the orders of the curves, left to right, that can be counted by the initial diagram stems. By virtue of the basic map derivation, points on the curves correspond to the chemical system equilibrium states. Unfortunately, interpretation of the system response in shifts against forces causes hardships in comparison to available experimental data: to the best of our knowledge, so far nobody looked at the problem of chemical equilibria from this point of view. That's why all comparisons so far may be only qualitative.

Majority of chemical reactions run at p,T=const; in this work we deal exclusively with isothermobaric chemical systems, and Gibbs' free energy is appropriate characteristic function. Some examples contain results of classical thermodynamic simulation, performed with computer simulation complexes either ASTRA-4 (ASTRA) [9] or HSC Chemistry for Windows (HSC) [10].

Finally, if appropriate we use term *clopen* instead of *non-isolated* system for brevity; in topology this is a blend of "closed *and* open", in DTd – "closed *or* open". Sometimes, on the euphony demand word "system" stands for subsystem, the real meaning is easily recognizable by context.

### 2.2. Definition of basic values.

When an isolated chemical system with one chemical reaction achieves TdE at its zero rate, reaction characteristic functions take on minimal values. *TdE is essentially the state of the chemical reaction.* There are additional conditions for equilibrium of clopen chemical system with its environment; we call such a state an open equilibrium (OpE), *it is the state of the chemical system.*

Standard change of Gibbs' free energy $\Delta G^0$ is a usual criterion of robustness for isolated reaction; we will employ instead more informative value, known in a slightly different than our definition from the law of simple proportions − *thermodynamic equivalent of chemical transformation*, η, a ratio between any participant amount, chemically transformed (i.e. not transported into or outside the system) along the way of isolated system *ab initio* to TdE (with asterisk), per its stoichiometric unit

(2) $$\eta_j = \Delta^* n_{kj}/\nu_{kj}.$$

K-participant amount of j-reaction is $\Delta^* n_{kj} = \eta_j \nu_{kj}$. The $\eta_j$ value depends on the system initial composition and the reaction $\Delta G^0$; this dependence is shown in Fig.1.

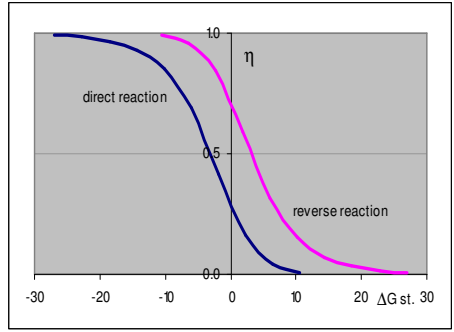

Fig.1. Thermodynamic equivalent of transformation $\eta_j$ vs. $\Delta G^0$, kJ·m$^{-1}$, direct reaction A+B=AB, reactants were taken in stoichiometric relations (HSC simulation).

Given stoichiometric equation, thermodynamic parameters (and therefore the reaction $\Delta G^0$), and the system elemental chemical composition, $\eta_j$ is the system invariant, unambiguously marking the state of TdE and taking on the same value for all reaction participants. This value may be found numerically via thermodynamic simulation of TdE in isolated system as a difference between the initial and the equilibrium amounts of a reaction participant. In DTd it serves as basic reference value for the states of chemical system.

Reaction coordinate $\xi_D$ was introduced by De Donder [11]

(3) $$d\xi_D = dn_{kj}/\nu_{kj}$$

with dimension of mole ($|\nu_{kj}|=0$!). We re-define it as

(4) $$d\xi_Z = (dn_{kj}/\nu_{kj})/\eta_j,$$

or

(5) $$d\xi_Z = dn_{kj}/\Delta^* n_{kj}.$$

In finite differences the reaction extent $\Delta\xi_Z$ is

(6) $$\Delta\xi_Z = \Delta n_{kj}/\Delta^* n_{kj},$$

$\Delta n_{kj}$ is the amount of k-moles transformed from the reaction initial to running states. So defined reaction extent is *a marker* of the system state regarding the reaction TdE with the initial value $\Delta\xi_Z=0$ and $\Delta\xi_Z=1$ at TdE. We define the system shift from TdE as

(7) $$\delta\xi_Z = 1 - \Delta\xi_Z.$$

We have $\delta\xi_Z > 0$ if reaction didn't reach TdE (or somehow was pushed back towards the initial state) and $\delta\xi_Z < 0$ if its state is beyond TdE. Obviously, $\delta\xi_Z = 1$ at the initial state, $\delta\xi_Z = 0$ is a specific characteristic of TdE, and in general $1 \geq \delta\xi_Z \geq 0$. In a figurative language, when reaction proceeds, its hosting system unfolds towards equilibrium. Positive shift means a forced compression of the unfolding system back to its initial state, while at negative shift the system state is forced towards the reactants exhaustion.

Obviously, defined this way reaction extent and the shift are dimensionless and measure the reaction accomplishment: *$\Delta\xi_Z$ is a distance from the initial state to running chemical equilibrium, reckoned in distances between the initial state and TdE*. A resembling restriction $0 \leq \xi_D \leq 1$ was discussed in a different context for the coordinate, defined by (3) [12] − there $\xi_D$ takes on unity only if at least one of the reactants is totally consumed. That means *logistic end* of reaction, and within this interval $\xi_D$ tallies the distance between the current state and the logistic end point. Such an approach has nothing to do with thermodynamics and equilibrium.

Reaction extent, based on De Donder's coordinate, may be designed in similarity with (6); it is linked to the newly defined value as

(8) $$\Delta\xi_D = \eta_j \Delta\xi_Z.$$

Further on we use exclusively z-subscribed values, omitting the subscript and retaining in writing only $\Delta_j$ and $\delta_j$ respectively. These quantities provide for a great convenience in equilibrium analysis: DTd employs $\eta_j$ as the basic theory parameter and $\delta_j$ is its basic variable. Introduced by de Donder reaction extent was important for development of thermodynamics, but has a great flaw due to its weird dimension of moles[I]; $\Delta\xi_Z$ fits the essence of generalized coordinate better.

The above introduced formulae allow for easy calculation of the component amounts. Indeed, initial amount of k-component in the system is $n^0_{kj}$, its transformed amount is $\Delta_j \eta_j \nu_{kj}$, the gain or loss are defined by the sign of stoichiometric coefficient, and in any state, defined by $\delta_j$, its equilibrium amount is $n_{kj} = n^0_{kj} + (1-\delta_j)\eta_j\nu_{kj}$. Now one can easily find mole fractions and their products for any subsystem. Although in some cases the distinction is only contextual, the DTd draws a clear line between the chemical reaction parameters and values and similar characteristics of the chemical system, harboring that reaction. Typical reaction parameters are $\nu_{kj}$, $\xi_Z$ and $\Delta G^0$. Besides $\eta_j$, Onsager coefficients $\alpha_{ij}$ with $i \neq j$, the extent $\Delta_j$ and the shift $\delta_j$ as well as $\Delta G$ exemplify the system parameters,

## 2.3. The Le Chatelier's response.

Changes in thermodynamic systems occur due to the action of *thermodynamic forces*, internal, that move system to equilibria, or external, that can act in arbitrary direction. The internal thermodynamic force in chemical systems was introduced by De Donder as thermodynamic affinity, a moving power of chemical transformations [11]

(9) $$A_j = -(\partial \Phi_j / \partial \xi_j)_{x,y},$$

$\Phi_j$ stands for major characteristic functions or enthalpy, *x* and *y* are appropriate external thermodynamic parameters. This definition matches general definition of force in physics as a negative derivative of potential by coordinate. Thermodynamic affinity in discrete form is

(10) $$A_j = -\Delta\Phi_j / \Delta\xi_j.$$

This value, the *eugenaffinity,* moves chemical reaction within isolated system to TdE, and is a typical reaction value. Accepting De Donder's definition of reaction extent, due to its weird dimension one immediately stumbles on a hard to explain dimension of affinity $[A_D] = J \cdot mol^{-2}$. Switching to dimensionless $\xi_Z$ turns the affinity dimension to the same as of the characteristic function.

Openness of any system is detectable through its interactions with environment; these interactions may be expressed via relationship between the external impact and the system response. We will formalize this relationship on the base of Le Chatelier's principle, defining the *Le Chatelier Response* (LCR) $\rho_j$ as a finite power series of the system shift from TdE, proportional to external thermodynamic force (TdF)

(11) $$\rho_j = \Sigma_{[0...\pi]} w_p \delta_j^p = (1/\alpha_j) F_{je},$$

$(1/\alpha_j)$ is proportionality coefficient (and $\pi$ is just an integer, not the Archimedes' constant!). The weights are unknown *a priori*, we don't have any samples for numerical evaluation of the weights, and we'd better off to get rid of them. Partly unfolding the sum in (11)

(12) $$w_0 + \Sigma_{[1...\pi]} w_p \delta_j^p = (1/\alpha_j) F_{je},$$

we find $w_0=0$ at $F_{je}=0$, and (11) becomes

(13) $$\Sigma_{[1...\pi]} w_p \delta_j^p = (1/\alpha_j) F_{je}.$$

---

[I] "The extent of reaction has the dimensions of moles. If $\xi$ increases from zero to one mole, *one mole of reaction* has occurred." [R. Mortimer. *Physical chemistry*, 2nd ed., p. 256. San Diego: Academic Press, (2000)]. I'll deeply appreciate if somebody can explain me what is *one mole of reaction*.

To eliminate the weights let's introduce an alternative expansion

(14) $$\omega_0(\delta_j) + \Sigma_{[1...\pi]}\delta_j^p = (1/\alpha_j)F_{je},$$

with such an auxiliary function $\omega_0(\delta_j)$, that

(15) $$\omega_0(\delta_j) = \Sigma_{[1...\pi]}(w_p-1)\delta_j^p.$$

Obviously, at $F_{je}=0$ and $\delta_j=0$ we have $\omega_0(0)=0$. Now, $\omega_0(\delta_j)<0$ at all $w_p<1$; because such a restriction is not enough grounded, we have a reason to suggest $\omega_0(\delta_j)>0$, which is also important for keeping positive the left hand side of (14). Because we simulate negative shifts via reverse reactions, $\Sigma_{[1...\pi]}\delta_j^p<0$ is not possible. This function increases with the $\delta_j$ and serves as adjustment parameter in the simulation code. Being an obvious palliative, $\omega_0(\delta_j)$ doesn't totally eliminate the uncertainty, but allows us to discover a variety of the chemical system states and paths.

Expression (14) accounts for non-linearity of the system response and leads to incomparably less bulky solutions than based on the alternative definition of the shift via power series of the external force [12]. In some cases variety of the system response modes depends on the system intricacy, and the limit value of power in the LCR $\pi$ may be loosely suggested a *system complexity parameter*.

Dimension of thermodynamic force is energy, $\delta_j$ is dimensionless, and dimension of $\alpha_j$ must be energy. Its value can be found from the experimental shift-force dependence, e.g. in experiments with electrochemical cells, where the external force is certainly known, but so far nobody has measured the shifts.

By the LCR approach we try "to deduce causes from effects" directly (Newton, *Optics*, quoted by [13]). Interestingly enough, that introducing this value we actually remove external thermodynamic force from the picture alike the elimination of force in general relativity theory, where "force is defined by the deviation of the particle from its "natural" path in space-time", and, wise versa, "the path of a particle … is called "natural" if no forces are exerted to the particle" [13]. By analogy with definition of the "natural" path as motion trajectory of free material system, perhaps first given by Hertz [14], we consider the path of isolated chemical system to equilibrium as its "natural" path.

**3. Premises of Discrete Thermodynamics.**

**3.1. Basic concepts.**

*A concept, demanding thermodynamic system to be divisible by parts in a way that those parts and their relations constitute the system itself, is one of the pillars of discrete thermodynamics of chemical equilibria.* The system equilibrium is achieved when all its subsystems are in equilibrium with each other, and each subsystem rests at its *open equilibrium, being connected to other subsystems via mutual interactions*. Actually, Guldberg and Vaage did the same, interpreting a system with simple chemical reaction as a *binary* system, whose only equilibrium is achieved when the rates of chemical transformations in its direct and reverse subsystems are equal.

Discrete thermodynamics defines chemical equilibria in a way, similar to used since long ago in mechanics, but, to the best of our knowledge, never before employed in chemical thermodynamics – as a balance point between the thermodynamic forces, acting within and against chemical system. These forces are thermodynamic affinity and resultant of external thermodynamic forces, which we will use mostly in the form of the LCR.

While classical theory is busy to describe mainly isolated systems, we take a close look at the system relations with its surrounding, which are essentially different for isolated, closed or open entities. Further on in this paper we deal mainly with closed systems, whose sources of external impact are abstract and represented by the action they produce (except the electrochemical systems).

**3.2. Derivation of the basic map.**

Onsager linear constitutive equations of non-equilibrium thermodynamics link the internal thermodynamic force $A_{ji}$ and the external forces $A_{je}$ with the reaction rate

(16) $$v_j = \alpha_{ji}A_{ji} + \Sigma\alpha_{je}A_{je}.$$

One may group all interactions within complex chemical system into dichotomial couples, each comprising a subsystem and its complement to the large, *mother* system. Denoting the resultant of all $A_{je}$ in each dichotomy as $\mathbf{A}_{je}=\Sigma A_{je}$ (and $\mathbf{o}_{je}\mathbf{A}_{je}$, as the second term in (16)), nullifying the rate $v_j=0$ and dividing (16) through by $\mathbf{o}_{ji}$ we obtain the sought condition of open system-environment equilibrium

(17) $\qquad\qquad\qquad\qquad A_{ji} + \mathbf{o}_j\mathbf{A}_{je} = 0,$

where $\mathbf{o}_j$ is dimensionless reduced Onsager's coefficient. Symbol $A_{ji}$ hints on the chemical nature of the internal force; external TdF may be of any nature, they act against the j-subsystem and their resultant must have opposite to $A_{ji}$ sign

(18) $\qquad\qquad\qquad\qquad A_{ji} - F_{je} = 0,$

$F_{je}=-\mathbf{o}_j\mathbf{A}_{je}$. The first term is the *bound affinity* [15], the j-subsystem internal thermodynamic force, caused by the system deviation from TdE and resistant to the external force, thus explaining why the bound affinity has opposite sign to the external force. The bound affinity is a system value: at TdF=0 (or at TdE) it doesn't exist. While equation (16) means the balance of fluxes, produced by the forces, equations (17) and (18) are explicitly displaying the balance of thermodynamic forces that constitute equilibrium between j-system and its complement.

Substituting $A_{ji}= -(\Delta\Phi_j/\Delta_j)_{x,y}$ in (18) and multiplying through by $(1-\delta_j)$ we obtain condition of chemical equilibrium as a logistic map

(19) $\qquad\qquad\qquad\qquad \Delta\Phi_j(\eta_j,\delta_j)_{x,y} + (1-\delta_j)F_{je} = 0.$

This is the basic idea of the discrete thermodynamics of chemical equilibria. Recalling the force value from (14), we get

(20) $\qquad\qquad\qquad\qquad \Delta\Phi_j(\eta_j,\delta_j)_{x,y} + \alpha_j[\omega_0(\delta_j)+\Sigma_{[1...\pi]}\delta_j^p](1-\delta_j) = 0.$

Map (20) includes the system shift from TdE under the external impact as the only variable given the pair of independent thermodynamic parameters, corresponding to chosen characteristic function. This result means that *in a clopen system, chemical equilibrium is achieved via competition between internal and external thermodynamic forces at the point of their consensus, where resultant of all thermodynamic forces equals to zero.* In clopen systems bound affinity appears as the system reaction to the external impact. Indeed, it is a real internal thermodynamic force, caused by external TdF and intended to return the system to TdE as the external TdF vanishes. The bound affinity manifests the subsystem resistance to external impact, it is the *system force of thermodynamic inertia*.

We have carried the above derivations keeping in mind open equilibrium as *equilibrium stationary state*. If thermodynamic forces, associated with non-zero flows are included into the sum of TdF, map (19) may as well describe non-equilibrium stationary states, not studied yet by DTd in details.

### 3.3. Chemical system map of states at p,T=const.

Substitution of $\Delta\Phi_j(\eta_j,\delta_j)_{x,y}$ in (20) by $\Delta G_j(\eta_j,\delta_j)$ leads to

(21) $\qquad\qquad\qquad\qquad \Delta G_j(\eta_j,\delta_j) + \alpha_j[\omega_0(\delta_j)+\Sigma_{[1...\pi]}\delta_j^p](1-\delta_j) = 0.$

Because $\Delta G_j(\eta_j,\delta_j)=\Delta G^0_j(\eta_j,0)+RT\ln\Pi_j(\eta_j,\delta_j)$, $\Delta G_j^0=-RT\ln K$, and $K=\Pi_j(\eta_j,0)$, and

(22) $\qquad\qquad\qquad\qquad \Delta G_j(\eta_j,\delta_j)=RT\ln[\Pi_j(\eta_j,\delta_j)/\Pi_j(\eta_j,0)].$

After substitution (22) into (21) and through division by $-RT$, map (21) becomes

(23) $\qquad\qquad\qquad\qquad \ln[\Pi_j(\eta_j,0)/\Pi_j(\eta_j,\delta_j)] - (\alpha_j/RT)[\omega_0(\delta_j)+\Sigma_{[1...\pi]}\delta_j^p](1-\delta_j) = 0.$

Because the $\alpha_j$ dimension is energy, one may interpret it as $\alpha_j=RT_{alt}$ with the *alternative* temperature, artificially introduced for the sake of logics: denoting $\tau_j=(\alpha_j/RT)=T_{alt}/T$, after a simple algebra at restricted $\pi$ (say, <20 or so) we arrive at the general map for isothermobaric chemical equilibrium

(24) $\qquad\qquad\qquad\qquad \ln[\Pi_j(\eta_j,0)/\Pi_j(\eta_j,\delta_j)] - \tau_j[\omega_0(\delta_j)(1-\delta_j)+ \delta_j(1-\delta_j^\pi)] = 0.$

Map (24) has a plenty in common with logistic maps, known in the theory of bio-populations [16,17]. The isothermobaric "growth" factor $\tau_j$ defines growth of the system deviation from TdE that drives chemical system into "far-from-equilibrium" area with typical complexity, bifurcations and chaotic behavior [16] (that's why in some previous publications we called this factor *reduced chaotic* *temperature*). In terms of Verhulst's model of bio-populations [18], its numerator $RT_{alt}$ represents external impact on the system ("demand for prey" [17]) while the denominator $RT$ is a measure of the system resistance to that impact. The growth factor plays critical role in evolution of chemical systems. Term $\Pi_j(\eta_{kj},0)/\Pi_j(\eta_{kj},\delta_j)$ is a ratio between conventional products of molar parts; its numerator represents "naturally" balanced population of isolated chemical system at TdE ($\delta_j=0$), and the denominator represents the externally disturbed population of clopen equilibrium system ($\delta_j\neq 0$).

Map (24) specifies conditions of chemical equilibrium *in any, isolated or clopen isothermobaric chemical system*. Its first term is change of the Gibbs' free energy, reduced by $RT$; it still keeps the classical form, slightly masked by presence of $\delta_j$, which makes the whole expression (24) uniform and clearly understandable. The second term reflects interactions between the j-subsystem and its environment and makes the difference between classical and DTd thermodynamics, causing a variety of "far-from-equilibrium" patterns. *Together they represent nothing else but full change of Gibbs' free energy in clopen chemical system, equal to zero when the system is at equilibrium with its environment.* At TdE, $\delta_j=0$ and the second term equals to zero; with the first term expressed in the classical manner, the map turns into classical change of Gibbs' free energy for isolated state, defining the state of TdE. The correspondence principle between classical and discrete thermodynamics of chemical equilibria is satisfied. In a sense, the second term is a binding energy of a subsystem with the mother system (a "chemical club membership fee").

We will use the term *map of states* for expression (19) and its particular forms (20) through (24); the reader will see that it makes more sense than just a substitution for the equation of state.

### 3.4. A corollary: constant of equilibrium as universal parameter.

Map (24) may be rewritten as

$$\Pi_j(\eta_j,0) = \Pi_j(\eta_j,\delta_j)\exp(\mu_j), \qquad (25)$$

where $\mu_j$ equals to the second term of (24). Now one can clearly see operational meaning of the map of states − it maps population of the isolated j-system at TdE onto the same system population at clopen equilibrium with $\delta_j\neq 0$: *states of the clopen chemical system can be deduced from its isolated state*. The left hand side of (25) is the reaction equilibrium constant $K_j$ by definition, and we arrive at a simple general rule for all equilibrium states including TdE in isothermobaric chemical systems

$$K_j = \Pi_j(\eta_j,\delta_j)\exp[\mu_j]. \qquad (26)$$

Given equilibrium constant and analytical or tabulated values of $\delta_j$ vs. $\tau_j$ one can calculate the chemical system composition at any $\delta_j$. In terms of the bio-populations theory, *equilibrium constant is a quantitative characteristic of the chemical system carrying capacity with regards to its population.*

In general, chemical system map of states is mapping its arbitrary state at $\delta_{j1}$ (e.g. $\delta_{j1}=0$) onto any other one with $\delta_{j2}\neq\delta_{j1}$. Both map (24) and expression (26) supposed to map the n-subscribed values onto (n+1); in our case both sides are subscribed identically. The reason is simple – our whole theory is built up around chemical equilibrium, where iterative calculations stop at the point where both values cannot be distinguished at given measure of accuracy, i.e. $\delta_{n+1}-\delta_n<\epsilon$. This is how the real system physically approaches its equilibrium with a measure $\epsilon$ of the system oscillations around it. As it follows from the derivation and structure of map (24), the open system behavior and states are governed by the resultant of the set of thermodynamic forces acting against it. Internal equilibrium, TdE, in a clopen system makes sense only as a reference state which it takes on being isolated from its environment or the mother system. As a result, the notion of detailed equilibrium, a hypothesis that traditionally plays key role in some approaches to complex chemical equilibria, should be replaced by the notion of the system equilibrium as a set of interrelated local binary equilibria, each defined for a certain subsystem with its own map of states (24).

### 4. Analysis of the Chemical System Response to External Impact.

### 4.1. Simple case of the Le Chatelier's response: linear force-shift dependence.

Let's consider a simplified linear LCR $\rho_j = \delta_j$ and

(27)  $\delta_j = (1/\alpha_j) F_{je.}$

Such a case occurred to be quite provable in *ad hoc* computer experiment. Imagine a chemical reaction between a metal oxide MeO and a reductant Re in isolated system; at TdE $\Delta_j=1$. Consider now a double oxide MeO·RO, where only MeO reacts with Re, RO is a binding, or restricting oxide, that lowers reacting activity of the MeO. Complex chemical system MeO-RO-Re comprises two open to each other subsystems with reactions MeO+Re=Me+ReO and MeO+RO=MeO·RO, both competing for possession of MeO. Now let MeO·RO react with Re. The outcome of the (MeO·RO+Re) reaction will be different than of (MeO +Re): TdF, originating from binding MeO into MeO·RO and acting against the subsystem MeO+Re, causes $\Delta_j<1$ and $\delta_j>0$ at complex chemical equilibrium. According to (15), this shift must be proportional to the TdF.

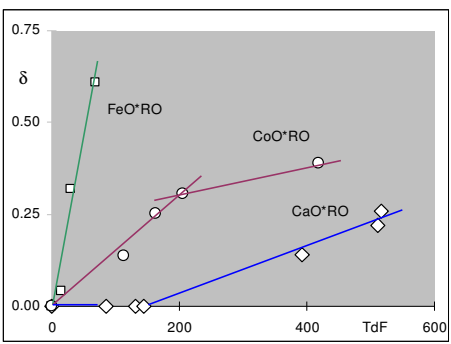

Fig.2. Linear LCR, system shift vs. TdF ($=-\Delta G_f^0/\Delta_j$), kJ·mol$^{-1}$, 298.15K, (MeO·RO+S) series, classical simulation (ASTRA-4).

Here $TdF_{je} = \Delta G^0_{MeO·RO}/\Delta_j$, a ratio between standard change of Gibbs' free energy of MeO·RO formation from the oxides and the (MeO+Re) reaction extent at complex (MeO·RO+Re) equilibrium. The data in Fig.2 were obtained by traditional thermodynamic simulations of the oxide interactions with sulfur, performed in two series – single oxides (MeO+S), MeO=FeO, CoO, CaO, to find $\eta_{MeO}$, and double oxides (MeO·RO+S), RO from the set {SiO2, Fe2O3, TiO2, WO3, Cr2O3}, to find $\eta_{MeO·RO}$. TdE states of the (MeO+S) systems naturally coincide with the reference scale zero point at $\delta_j=0$; shifts of the (MeO·RO+S) systems were calculated as

(28)  $\delta_j = 1 - \Delta n_{MeO·RO}/\eta_{MeO}$,

$\Delta n_{MeO·RO}$ is the number of MeO moles, transformed to equilibrium in the (MeO·RO+S) reaction. Results in Fig.2 confirm the shift vs. TdF linearity in the particular investigated systems and justify usage of the LCR (27) as the first approximation.

**4.2. Graphic solutions to the isothermobaric chemical system map of states.**

Map (24) is transcendental: its solution demands the numerical methods. Graphical images to the basic map solutions are specific mono-bifurcation pitchfork bifurcation diagrams, partly resembling the diagrams known from the bio-populations theory [16], but having several unknown before features. Set of such diagrams for direct and reverse reactions in two chemical systems with different stoichiometry and varying $\eta_j$ are shown in Fig.3,4, the reactants were in stoichiometric proportions.

All diagrams for map (240 are in general the same and have three distinctive areas, present in both direct and reverse parts; each area has specific meaning for the chemical system. The area with $\delta_j=0$, the TdE area arrives first. The second is the area of open equilibrium, OpE, with $\delta_j \neq 0$, where the map still has only one solution and the system state changes reversibly; here the system states are sitting on classical thermodynamic branch. When the branch becomes unstable, all diagrams (besides corresponding to $\eta_j \approx 1$) experience unique bifurcation of period 2 at the limit point $\tau_{OpELim}$ followed by the area of the bi-stable states with duplicate $\delta_j$ values. Bifurcation diagrams as graphic solutions to the basic map are the *chemical system diagrams of states*, covering all conceivable states from TdE back to the initial state with $\delta_j=1$ on the upper branch, where the system may be forced to return to.

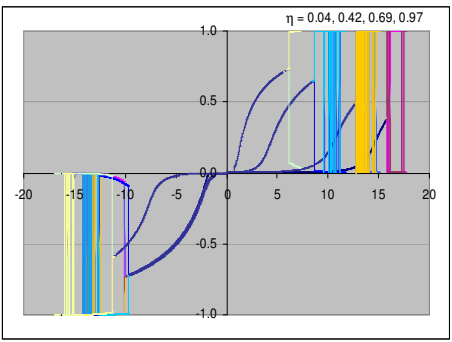

Fig.3. Bifurcation diagrams for the system with direct reaction A+B=AB, $\pi=1$.

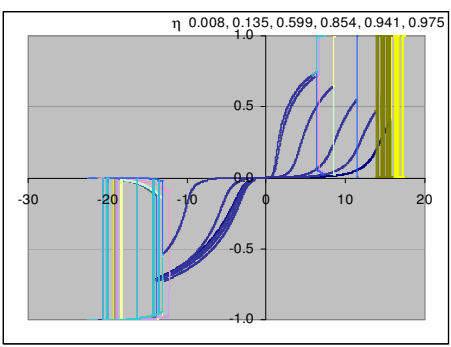

Fig.4. Bifurcation diagrams for the system with direct reaction $3A+B=A_3B$, $\pi=1$.

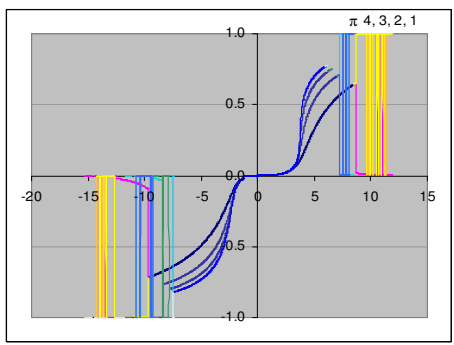

Fig.5. Bifurcation diagrams for the system with direct reaction A+B=AB, $\eta_j=0.687$, varying $\pi$.

Bifurcation diagrams with varying $\pi$ are shown in Fig.5. Their shapes are similar to the diagrams in previous pictures; there are no bifurcations in direct reactions with $\pi=3$ and $\pi=4$. We have only vague idea of whether the complexity factor plays essential role in chemical system evolution.

The above discussed bifurcation diagrams give a good idea of what happens to the system with increase of the control parameter $\tau_j$. While this parameter has a good enough explanation in both bio-populations and the chemical systems, unfortunately it is a not-measurable value and there is no way to find the $\delta_j$–$\tau_j$ relationship experimentally. That's why along with the *static* bifurcation diagrams $\delta_j$ vs. $\tau_j$ we also consider *dynamic* diagrams $\delta_j$ vs. TdF, as it was done for the electrochemical systems described below. We expect future experimental data to represent namely the shift-force dependence, because thermodynamic forces are either measurable or computable from direct experimental data.

Within the already developed DTd formalism we can without compromises reconstruct the force from the $\delta_j$ and $\tau_j$ values only in the TdE and OpE area; the shapes of these diagrams repeat in a different scale the shapes of the static diagrams in those areas, Fig.6. We don't know yet how to make a unilateral reconstruction beyond bifurcation point.

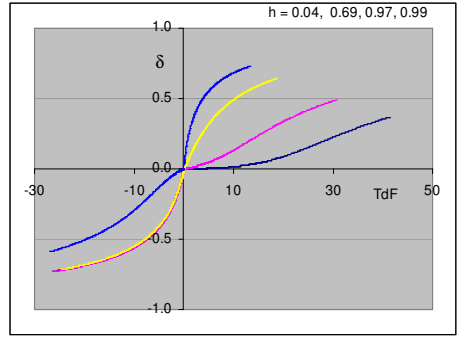

Fig.6. System shift vs. TdF, kJ·m$^{-1}$, system with reaction A+B=AB for various $\eta_j$, $\pi$=1.

**4.3. The chemical system domain of states.**

Varying initial chemical composition, $\Delta G^0_j$ and $\pi$, we obtain a continuous set of the state diagrams for a system with certain chemical reaction. The set constitutes the *chemical system domain of* states, which can be designed for any individual reaction with widely varied parameters and fills in entirely the first and the third quadrants in the flat case or appropriate cubes in 3-d Cartesian coordinates.

The least expected and very interesting result of the new theory is extensive TdE area in clopen systems within a wide range of parameters, which contracts to the zeroth point of the reference frame in isolated systems. It is natural to expect, that the more robust is the reaction, the more expressed must be its resistance to external impact. This statement is well illustrated by Fig.3,4 – the larger is the $\eta_j$ (or the more negative is the reaction $\Delta G^0_j$), the longer is the TdE area. The states along this area are not sensitive to changes of $\tau_j$ (or external TdF), thus representing a zone of indifferent equilibrium, currently rarely referred [19]. One can see a similar delay in Fig.2 on the $\delta_j$ vs. TdF curve for CaO·RO series. At the same time, $\tau_{TdELim}$ is the *classical limit* of the chemical system response to external impact; below this limit chemical system reacts in the classical pattern. In computer experiments within a wide range of reaction stoichiometry and initial compositions, two features of the classical limit were observed (mainly at $\omega_0(\delta_j)$=0, qualitatively giving the same results):

- *when $\eta_j \to 0$, the classical limit tends to the sum of stoichiometric coefficients of the reaction products: $\tau_{TdELim|\eta_j \to 0} \to \Sigma \nu_{pj}$.*
- *regardless the $\Sigma \nu_{pj}$ value, the larger is $\eta_j$ the farther is the classical limit from zero along abscissa; at $\eta_j \to 1$ the classical limit tends to infinity: $\tau_{TdELim|\eta_j \to 1} \to \infty$. This is the case of absolute irreversibility.*

The first observation is relevant to the systems with thermodynamically not-robust reactions, the second – to the systems with robust reactions. The second observation means that the more robust is the reaction, the more powerful should be the external impact to move the system off the classical limit. *This observation turns the thermodynamic equivalent of transformation $\eta_j$ to an effective criterion of chemical irreversibility*. One can find more details in [20]. Interestingly, that the $\tau_{TdELim}$ value may be calculated analytically showing a good match with simulated results [21].

The OpE area represents a watershed between the point attractor, whose basin is probabilistically frozen kingdom of classical chemical thermodynamics at TdE, and the strange attractor, the "far-from-equilibrium" wild republic of the states-by-chance beyond the OpE limit. Position of the bifurcation point on abscissa exactly defines how far is the "far-from-equilibrium" and $\tau_{OpELim}$ also unilaterally depends on $\eta_j$: the larger is $\eta_j$ the further from the TdE is bifurcation point, i.e. the stronger is the system resistance against bifurcations.

As $\tau_j$ and TdF increase further, bifurcation area stays either unchanged or experience instabilities, appearing as series of chaotic oscillations instead of the period dubbing, typical for the habitual bifurcation diagrams. They are clearly expressed on many bifurcation diagrams and have been never predicted by classical thermodynamics.

Domain of states as a whole is located along positive $\delta$–axis up to $\delta_j=0$ for no one chemical reaction cannot be pushed back below its initial state; for $\delta_j<0$, when reaction is promoted to the side of products, its lower limit is defined by the reactants ratio and coincides with the reaction logistic limit.

Due to the above redefined basic thermodynamic values and followed then derivations we have set a new reference frame, where the home state of isolated system, TdE, corresponds to the zeroth point; ordinate of the frame is the system deviation from TdE, and abscissa is either $\tau_j$ or TdF. Dynamic diagrams constitute the chemical system *dynamic domain of states*.

Another feature of the domains of states is their fractal properties. The size of domain along the $\tau_j$ axis is reversely proportional to the simulation parameters t and $\alpha$ (see Appendix A). Results of simulation are shown in Fig.8; as it is easy to see, the whole bifurcation diagram shrinks toward the zeroth point of the reference system as the $\alpha$ value increases.

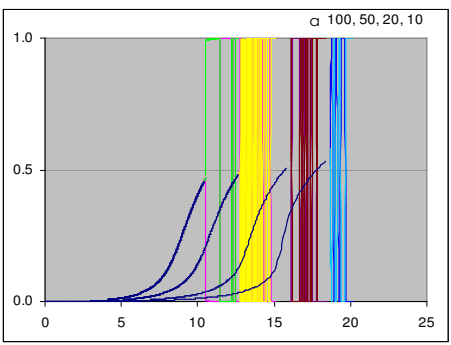

Fig.7. Fractal properties of the chemical system domain of states, A+B=AB, $\eta_j=0.687$, $\pi=1$, Inverse "iteration stick" lengths (see Appendix A) are shown on the upper field.

**4.4. Thermodynamically predicted chemical oscillations.**

Bifurcations and chemical oscillations belong to the most interesting phenomena observed in open chemical systems. The relevant boom in science is well back in time, but those events are still observable in laboratories and in Nature. We have found that as the external impact strengthens and external TdF takes on a certain magnitude, the chemical systems experience natural *chaotic* oscillations between the branches of bifurcation area. They are distinctively visible in Fig.3-5 and Fig.7; rescaled fragment of the oscillation spectrum is shown in Fig.8. As it follows from our data, predicted by DTd chemical oscillations exist within a restricted range of external forces, which is in agreement with experimental data for oscillating reactions [16] as well as with the data for electrochemical systems (experimental and in Chapter 6 of this paper). Increase of $\eta_j$ leads to more ostensible oscillations with shifting locations of the oscillations zone in the bifurcation area. The oscillations are short-time inversions of the subsystem states.

The above found oscillations occur in closed chemical systems with independent sources of external force. To analyze the systems with auto-oscillations like the Bray-Liebhafsky [22,23] or famous BZ [24] reactions, one has to consider simultaneously all parts of the chemical system as open systems as it was explained earlier. Although intuitively it may be uncomfortable to believe, that the oscillations exist in such a simple system as A+B=AB, please pay attention to how simple are the reactions, studied by Bray and Liebhafsky, or in some electrochemical systems. Observed in this work chemical oscillations follow naturally from DTd and feature high repeatability, but their spectra in many cases are far not simple. They are so peculiar for different systems and so sensitive to their parameters, that there must certainly be a dependence of the spectrum structure upon the said factors. These oscillations are predictable by DTd without a trace of kinetics or autocatalytic considerations.

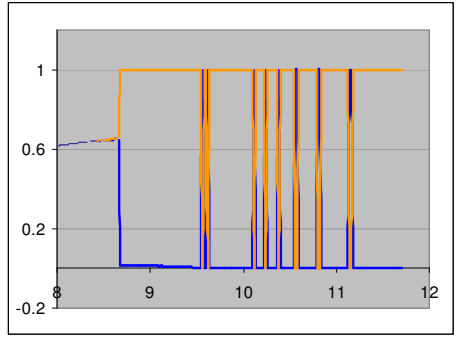

Fig.8. Rescaled fragment of the diagram with chemical oscillations, system with direct reaction A+B=AB, η=0.687, π=1.

## 5. Practical Use of Discrete Thermodynamics.

### 5.1. Discrete thermodynamics and activity coefficients.

The role of thermodynamic activity coefficients in the classical theory was briefly mentioned above. If deduced from probabilistic considerations, they could be more organically woven into the classical canvas of based on probabilities thermodynamics. Indeed, if only one reaction runs in a system, the outcome is defined by a probability of the reaction participants to collide, or by the product of their *a priori* probabilities to occur simultaneously in a certain point of the reaction space. These probabilities equal to their mole fractions in the reacting mixture, and we arrive at traditional mass action law. The situation gets worse if several coupled chemical reactions are running simultaneously in the system. Let $\{R_{k1}, R_{k2}, R_{k3}, \ldots\}$ be a set of the only possible events on the reaction space $S_j$, competing for the k-component. Because its amount in the system is restricted, the outcomes of events are mutually dependent, and now the mass action law contains *conditional* probabilities. Let event $A_k$ is occurrence of any $R_{kj}$ on the space $S_j$; their probabilities are conditional $p(R_{kj}|A_k)$ and *a priori* $p(R_{kj})$. Then the statement known as Bayes' theorem [25]

(29) $\quad p(R_{ki}|A_k) = p(R_{kj}) p(A_k|R_{ki}) / \Sigma[p(R_{kj}) p(A_k|R_{kj})].$

defines conditional probability of $R_{ki}$ given $A_k$. Ratio $\gamma_{kj}^{v_{kj}} = p(R_{ki}|A_k)/p(R_{ki})$ is the probabilistic activity coefficient; this is exactly the coefficient of thermodynamic activity, introduced by G. Lewis in a different form. Combination of the coefficients, related to components of the same reaction, or to the dwellers of the same subsystem, designed exactly as appropriate mole fractions product, form excessive thermodynamic functions, represented by the third term in following expression

(30) $\quad \Delta G_j = \Delta G^0_j + RT\ln\Pi_j(x_{ki}^{v_{kj}})] + RT\ln\Pi(\gamma_{ki}^{v_{kj}}).$

Though the Bayes' theorem weighs prior information with empirical evidence, it is still the best tool to demonstrate the probabilistic meaning of activity coefficients. Comparison between map (24) and equation (30) with $\Pi_j(\eta_j, \delta_j) = \Pi_j(x_{ki}^{v\ kj})$ prompts us to suggest the second term in (24) and the third term in (30) to have the same meaning, both reflect the system clopenness and external impact. Supposing for simplicity $\omega_0(\delta_j)=0$, π=1 and reducing the third term of (30) by RT, we arrive at

(31) $\quad \tau_j \delta_j (1-\delta_j) = -\ln\Pi(\gamma_{ki}^{v_{kj}}).$

With the only one activity coefficient per subsystem we have

(32) $\quad \delta_j = (1/\tau_j)[(-\ln\gamma_{ki})/(1-\delta_j)].$

To validate this expression, we carried out classical simulation (HSC) of reaction

(33) $\quad 2CoO + 4S = CoS_2 + CoS + SO_2$

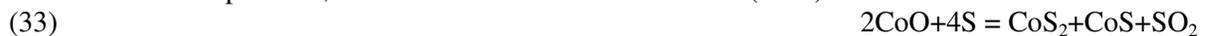

at p=0.1 Pa, T=1000K with reactants, taken in stoichiometric ratio, varying activity coefficients of CoO. For comparison we have simulated corresponding bifurcation diagrams; obviously, reduced activity of a reactant cuts off the reaction outcome and $\delta_j>0$. External TdF at equilibrium was counted in two ways – first as $F_{ej}=\ln[\Pi_j(\eta_j,0)/\Pi_j(\eta_j,\delta_j)]/(1-\delta_j)$ and then as $F_\gamma=(-\ln\gamma_{kj})/(1-\delta_j)$. Obtained by two ways force-shift dependences for this reaction are shown in Fig.9. The above found similarity and equation (31) form a ground for independent method to find out coefficients of thermodynamic activity for various applications. Though the coefficients are out of the DTD concept, using them in some cases is still simpler and so far inevitable. From this point of view, what we have done in DTD with regard to activity coefficients may serve as a fresh view at their nature and as an alternative method to calculate them. More details regarding the method may be found in [20].

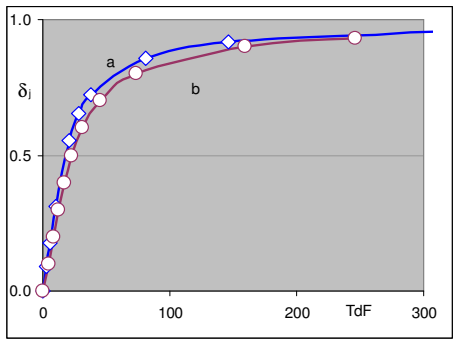

Fig.9. Shift – TdF dependences, reaction (33), 1000K, curves are relevant to dimensionless TdF: a - as $F_{ej}$, b - as $F_\gamma$.

Actual link between two groups of results is stronger than mere numerical coincidence. Here we encounter a similarity between the ideal/non-ideal behavior of chemical system, that affects (or is expressed via) thermodynamic activities of certain chemical species, belonging to the system. Chemical reaction in an isolated system is able to achieve TdE state with $\Delta^*_j=1$, while in a clopen system $\Delta^*_j\neq 1$. Actually $\Delta^*_j$ in a sense plays role of the *chemical system coefficient of thermodynamic ideality*, affecting all species of the system via their activity coefficients.

**5.2. Simple example of discrete thermodynamic simulation.**

A conditional reflex, deeply rooted in the mentality of those, who deal with equilibria, prompts them to identify automatically any chemical equilibrium as TdE. This habit is heavily supported by all current computer software for thermodynamic equilibrium simulation, which is based on this approach for it is simple and convenient. But not true. It provides for more or less realistic quantitative analysis only for a few systems, hosting reactions with big enough negative $\Delta G^0$. Discrete thermodynamics changes the situation, opening new opportunities for advanced thermodynamic simulation and analysis of real chemical systems in a wide range of $\Delta G^0$.

Based on map (24), thermodynamic simulation of complex chemical equilibria takes into account their nature as comprising mutually interacting clopen entities with their states fully determined by the shift from TdE. Domain of states for each entity is totally individual, independent and is known beforehand (we design and create it by ourselves with described in Appendix A technique or alike). Subsystem interaction with its complement defines state of the subsystem, a point in its TdE or OpE areas with unilaterally linked together $\tau_j$ and $\delta_j$ or two points in the bifurcation area. Joint solution of the maps (24) for all subsystems allows us to find their amounts at complex equilibrium.

To show the numbers and to illustrate how it works in the simplest case, we have calculated equilibria in mixtures with double oxides along the previously described in this paper path. $\Delta G^0$ of the double-oxide formation from single oxides was used to find corresponding $\tau_j$ values. This is the chain of calculations: $\Delta G^0_{f(\text{MeO·RO})}$ was considered the excess function for any of the two single oxides; it gave us the $\ln\gamma_j$ from the third term of (30) ; then $\tau_j\delta_j$ was obtained as $(-\ln\gamma_{kj})/(1-\delta_j)$, and finally the bifurcation diagram was used to get the appropriate $\tau_j$ value. Equilibria in homological reaction series
(34)  $\quad\quad\quad\quad\quad\quad\quad\quad\quad\quad\quad\quad\quad\quad\quad\quad\quad\quad\quad\quad 2CoO·RO+4S=CoS_2+CoS+SO_2+RO$
were simulated at p=0.1 Pa, T=1,000K with reactants taken in stoichiometric ratio, R={$TiO_2,WO_3,Cr_2O_3$}. To compare the results, classical thermodynamic simulation was performed on reaction (33) with variable CoO activity coefficients using HSC [10], $(-RT\ln\gamma_j)$ was considered the excess function. Simulation results are shown in Table II. The DTd data are the graphical solutions to corresponding maps (24); their logic is clear from Fig.10. The curves are plotted in coordinates $\Delta^*_j$ vs. either $\Delta G_j/RT=-\ln[\Pi_j(\eta_j,0)/\Pi_j(\eta_j,\delta_j)]$ (ascending curves) for reactions of double oxide formation from oxides or vs. the term $\tau_j\delta_j(1-\delta_j)$ (descending curve). Intersections of the rising curves with the descending curve give the numerical values of corresponding equilibrium reaction coordinates. Expected value of $\Delta^*_j=1$ (or $\delta_j=0$) for free CoO is confirmed in a triple match point, where the ascending CoO curve hits the ordinate and meets the descending curve, thus marking undisturbed TdE. The major point to pay attention to is a well pronounced and increasing with certain regularity difference between the "classical" and DTd reaction shifts in Table I. As it was said, a quite obvious conclusion is that *discrete thermodynamic equilibrium simulation in some cases allows us to avoid usage of thermodynamic activity coefficients.* That extends the advantage of discrete thermodynamic simulation even further because activity coefficients are expensive and normally should be found before the simulation.

Table I. Equilibrium reaction extents and shifts in reaction (34) homological series.

| Reactant | CoO | CoO·$TiO_2$ | CoO·$WO_3$ | CoO·$Cr_2O_3$ |
|---|---|---|---|---|
| $(-\Delta G^0_{f(CoO·RO)}/RT)$, kJ·m$^{-1}$ | 0.00 | 3.77 | 6.17 | 7.2 |
| $\Delta$ simulated, HSC | 1.00 | 0.92 | 0.88 | 0.86 |
| $\Delta$ graphical, DTd | 1.00 | 0.9 | 0.82 | 0.77 |
| $\delta$ simulated, HSC | 0.00 | 0.08 | 0.12 | 0.14 |
| $\delta$ graphical, DTd | 0.00 | 0.10 | 0.18 | 0.23 |

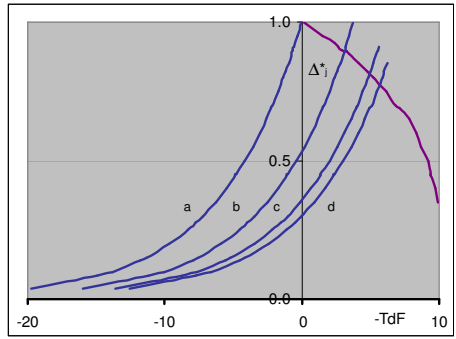

Fig.10. Equilibrium reaction extent $\Delta^*_j$ vs. $(-TdF)$, kJ·m$^{-1}$. Reaction (34), T=1000 K, ascending curves: a−CoO, b,c,d−CoO·RO, RO=$TiO_2$, $Cr_2O_3$, $WO_3$; descending − the second term of map (24).

## 6. Applications with Explicit Thermodynamic Forces: the Electrochemical Systems.

### 6.1. Electrochemical system map of states.

Population of electrochemical system and interactions between its parts and with its environment are essentially different from regular chemical systems due to presence of charged particles, redox reactions, electrical potentials/electromotive forces, and electric currents. At the DTd approach, the electrochemical cell comprises two subsystems − the chemical and the electrochemical (actually, electrical); both are functionally inseparable, their states change strongly interdependently. In conventional thermodynamic glossary both are the closed systems with no material exchange.

We will give a logical, conceptual derivation of the relationship between the chemical subsystem shift from TdF and the electrode potentials; more detailed derivation may be found in [26]. Suppose chemical subsystem of the electrochemical cell without submerged electrodes at TdE, its $\Delta G_{jc}=0$, subscript c stands for "chemical". Now we submerge electrodes, a working and a standard, into the electrolyte and new equilibrium, this time between the chemical and the electrochemical parts is settled up. The observer will find a potential on the working electrode, $E_j^*$, the chemical subsystem state is changed, now $\Delta G_{jc}^*=0$ and $\delta_{jc}^* \neq 0$, asterisk refers values to electrochemical equilibrium. The $E_j^*$ value is the measure of the electrical and $\delta_{jc}^*$ is the measure of the chemical changes. Electrochemical equilibrium is complex equilibrium between mutually balanced subsystems of the cell. Recalling the expression for electrical force, corresponding to the electrode potential $E_j$ as $n_j\mathcal{F}E_j$, with the amount of transferred electron charges $n_j$ and Faraday number $\mathcal{F}$, or $n_j\mathcal{F}E_j^*$ at equilibrium, from the balance between internal and external forces we obtain the map for open circuit

(35) $$\Delta G_{jc}^* - (1-\delta_{jc}^*)n_j\mathcal{F}E_j^* = 0.$$

It differs from the Nernst equation for equilibrium electrode potential by factor $(1-\delta_{jc}^*)$ [27]. The reason is in the choice of reference state – we use TdE and put $\delta_{jc}^* \neq 0$ at electrochemical equilibrium, while in the Nernst approach the electrochemical equilibrium, deviation from which is zero, was taken as the reference state.

If there is a voltage difference between the cell electrodes, but electrical current in a closed circuit is slow, the state of the chemical subsystem is changing slowly, the states of both subsystems change synchronously, the cell still stays in electrochemical equilibrium, and general map for electrochemical equilibrium is

(36) $$\Delta G_{jc} - (1-\delta_{jc})n_j\mathcal{F}\Delta\varphi_j = 0,$$

now the formula contains running values (without asterisk), and taking into account (22) we arrive at

(37) $$\ln[\Pi(\eta_j,0_j)/\Pi(\eta_j,\delta_{jc})] + (1-\delta_{jc})n_j\mathcal{F}\Delta\varphi_j/RT = 0.$$

This map reflects the fact that if the chemical system hasn't achieved its TdE (which is barely possible with electrodes), the $(1-\delta_{jc})$ factor may be considered an electron transfer coefficient.

**6.2. Electrochemical system domain of states and oscillations.**

Factor $(1-\delta_{jc})$ in (37) provides for feedback, turning electrochemical cell into dynamic system; its graphical solutions are the same fork bifurcation diagrams as in chemical systems. Fig.11 shows combined simulation results in $\delta_{jc}-\Delta\varphi_j$ coordinates – one for map (37) and another one, relevant to quasi-classical simulation described above, Fig.12 – a rescaled fragment of the oscillation spectrum. Electrochemical bifurcation diagrams have an ascending stem, growing up from the zeroth point, and then experience bifurcation period 2; within certain ranges of parameters they also experience well pronounced oscillations between the upper and the lower bifurcation branches as the external force changes. The oscillation spectra are located within restricted intervals of the external TdF, and their lines and line groups are separated by clearly visible windows of stability. We found that every set of the parameters in (37) has its own influence on the oscillation spectrum signature; changing them one can create a full set of bifurcation diagrams, constituting the electrochemical system *domain of states* and entirely filling in the I and the III quadrants of coordinate plane ($\delta_j$ and $\Delta\varphi_j$ both may take on plus and minus signs independently). Electrochemical oscillations occur within narrow areas, restricted by certain values of $\Delta G_j^0$ and the cell potential; this feature for a simple redox reaction A+B=AB (say, A−ne⁻=A$^{n+}$, B+ne⁻=B$^{n-}$) is exemplified in Fig.13. Oscillations occur within the cusped bodies. Map (37) describes equilibrium steady state in electrochemical cells.

The envelopes of the bifurcation areas are different for the diagrams obtained with LCR and obtained with the explicit force. The reason could be in $\omega_0(\delta_j)$ – we have observed that the smaller is its value, the more curved and smooth are the initial parts of bifurcation branches. Experimental oscillation spectra show a big variety of the envelope curve shapes including all ours.

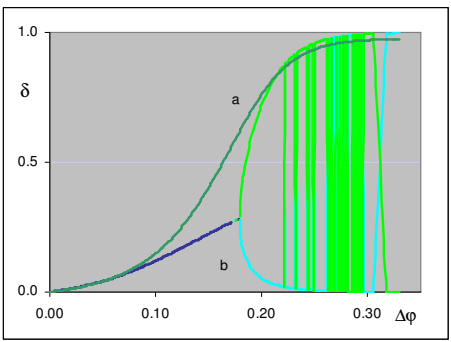

Fig.11. Comparison btw "classical" (a) and DTd (b) shift dependence upon cell potential. Redox reaction A+B=AB, $n_j=1$, $\Delta G^0_j = -18.0$ kJ·m$^{-1}$, T=293.15K, $|\Delta\varphi|$=V.

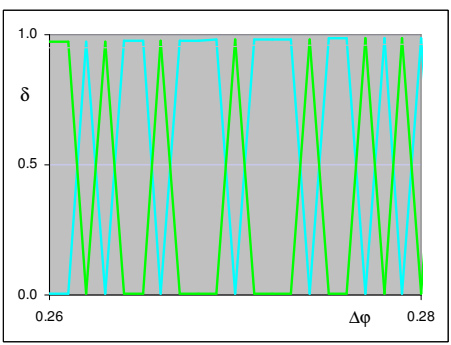

Fig.12. Rescaled fragment of the oscillation spectrum from Fig.11.

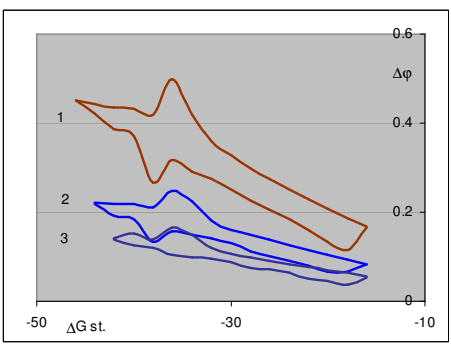

Fig.13. Areas of electrochemical oscillations for different numbers of transferred electrons (numbers at the curves), reaction A+B=AB, T=293.15K, $|\Delta G^0|$=kJ·m$^{-1}$, $|\Delta\varphi|$=V.

### 6.3. Experimental data vs. simulated electrochemical oscillations.

Our results were plotted in $\delta_{jc}$–$\Delta\varphi$ graphs while experimental points are usually presented in $I$–$\Delta\varphi$ coordinates. Electrical current within the electrolyte is a flow of charged particles between the areas with different potentials, or with different $\delta_{jc}$. There must be a correlation between the electrical current value and difference in the chemical subsystem states; we certainly don't know it yet, and we took a look only for a similarity between our results and experimental data. Experimental graphs, adopted from two publications, are shown in Fig.14 and Fig.15. The authors of [28] have mentioned that the origin of the oscillatory behavior of metals in acidic electrolyte (Fig.13) was not well understood and they supposed that the oscillations were caused by periodic film formation and dissolution on the electrode surface as well as periodic changes of the film chemical contents. As it was found in [28], the oscillations occurred within narrow range of the potentials for investigated concentrations of the trichloroacetic acid solutions.

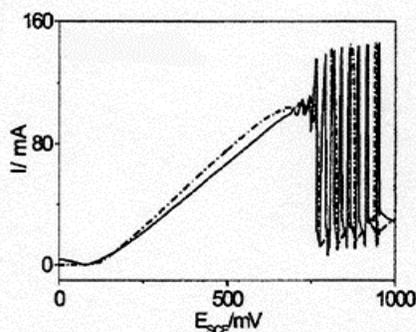

Fig.14. Polarization behavior of the Cu/CCl$_3$COOH system; concentration of CCl$_3$COOH − 1 M·dm$^{-3}$, adopted from [28].

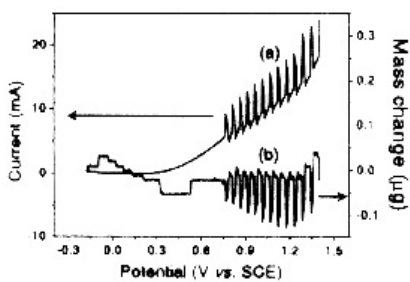

Fig.15. Linear sweep voltammogram at a platinized Pt electrode, 0.1 M H$_2$SO$_4$+1 M HCHO, a) current-potential curve, b) EQCM mass change curve, adopted from [29].

Following similarities between the thermodynamically predicted and experimentally observed electrochemical oscillations may be noticed:
- oscillation areas are located within narrow voltage intervals;
- oscillation spectra experience windows without oscillations;
- in many cases experimental magnitude of oscillations is restricted by lines similar to lower and upper bifurcation branches of the simulated bifurcation diagrams.

### 7. Conclusion: Understanding of Discrete Thermodynamics

The reader, saying that there were more problems set than solved in this work, is not very wrong. But the above results would be purely scholastic, if there were no physico-chemical contents in the maps we have derived. Besides the Le Chatelier response, whose concept originated from quite usual in physics approaches and was modified in this work to account for shift-force relationship in chemical systems, everything else in this work was obtained directly from the recognized ideas of contemporary thermodynamics. Excluding loosely defined value of $\omega_0(\delta_j)$, the basic map of states has been derived, not postulated or composed, like e.g. Schrődinger's equation in quantum mechanics.

### 7.1. The origins of discrete thermodynamics.

Curiously enough, the DTD concept of binary equilibria and dichotomy treatment of complex chemical systems as the balance of forces between its subsystems was inspired by description of the arch by Leonardo da Vinci: "*The arch is nothing else than a force originated by two weaknesses, …* *as the arch is a composite force, it remains in equilibrium because the thrust is equal from both sides.*" [30]. The same is valid for any complex system, including chemical – complex equilibrium is the equilibrium between its subsystems, originated from and supported by their interactions. At this point we recall famous discovery of Aristotle – "*The whole is more than the sum of its parts*".

When the author was studying theoretical mechanics during the remote from now years of tuition, he was strikingly impressed by d'Alembert's principle. Quoting C. Lanczos [31], "*With a stroke of genius eminent French mathematician and philosopher d'Alembert succeeded in extending the applicability of the principle of the virtual work from statics to dynamics*". By adding forces of inertia to the Newton's equations, he created the method to solve dynamic problems as static. The idea of equation (18) and basic maps of states of chemical systems with the bound affinity, a value originated in the system move from TdE as the force of thermodynamic inertia, is the mould of d'Alembert's principle. DTd offers the way to solve classically non-equilibrium problems as equilibrium ones. Thermodynamic version of the principle may be worded as: *any state of chemical system may be considered equilibrium, if thermodynamic forces of inertia are added to the external thermodynamic forces.* More details and some other important corollaries of the principle one can find in [32].

**7.2. Discrete thermodynamics today.**

One can anticipate two potential readers' questions: Why all changes are in $\Delta s$? Why $\Delta \xi_z = 1$ at equilibrium? First, the well known fact is that chances to find the sought variety of states using derivatives and differentials are very slim, if not none. On the other hand, the readers of this paper most probably are familiar with traditional thermodynamic $\Delta-$thinking. Thermodynamic affinity in the finite differences is one of the important expressions of the theory, and satisfaction of the equality

(38) $\qquad A_j = -\partial G_j / \partial \xi_j = -\Delta G_j / \Delta \xi_j$

is a must. Its validity can be easily proved for the species formation from elements (the major reaction unit in thermodynamic simulation), and then, using the Hess' law of constant heat summation [33] may be extended for chemical reaction of any stoichiometry.

As concerns to $\Delta_z$, one may have recognized that the reaction extent as it was introduced by De Donder and as we mentioned earlier turned out practically not useful. Linking the reaction extent value to the system states as a strong numerical marker allows us to follow the chemical system evolution from its initial state to TdE. At this point $\Delta_z = 1$ looks the most natural and simple; resemblance between the system and the species ideality was discussed in subchapter 5.1.

The old truth about thermodynamics is that the more detailed and appropriate is the system thermodynamic description, the more effective will be its thermodynamic analysis. Due to such innovations like new for chemical thermodynamics approach to chemical equilibrium as balance of thermodynamic forces; introduction of the system shift from TdE in relation to the reaction extent; the system state evaluation via the shift; introduction of the thermodynamic equivalent of chemical transformation; amended definition of thermodynamic affinity; and normalized to constants the values of reaction extent and reaction shift at TdE, discrete thermodynamics is able to show and describe such unknown for classical theory phenomena as open equilibria, bifurcations, bi-stability and chemical oscillations. DTd introduced a new, previously unknown transcendent logistic map of states of chemical systems, which reflects instabilities in the systems as chaotic oscillations instead of dubbing the bifurcation period; on its solutions is based such a new notion as domain of states of the chemical system. DTd clearly presents chemical equilibrium as a system phenomenon.

DTd intent is to extend horizons of our understanding and to solve previously insoluble problems in real systems. Based on DTd new method to simulate complex chemical equilibria will definitely bring new opportunities to thermodynamic analysis.

Thermodynamically predicted chaotic oscillations in chemical and electrochemical systems, discovered in this work, are very important for understanding and analysis of the open system behavior. So far all oscillation phenomena were explained exclusively on the kinetic basis. This is how series of well known models like "brusselator", "oregonator", etc. and all explanations of the famous chemical oscillating reactions were born. Title of one publication – "Chemical oscillations arise solely from kinetic nonlinearity and hence can occur near equilibrium" [34] (sounding as if the authors are trying to convince themselves) – may serve as a manifesto of that approach, and kinetic perception of chemical oscillations reigns in all publications and basic monographs. We do not deny ability of kinetic models to produce images, close enough to experimentally observed chemical oscillations; we just offer alternative point of view and explanation of the oscillations origin and progress, which is much simpler than the kinetic approach. Future will find a balance between them.

As a praise of DTd one should mention simplicity, laconism and clarity of the basic ideas and derivations, ability to cover on a unique basis more experimental facts than any other of the current theories, and compliance with the correspondence principle. Although so far the qualitative DTd results are prevailing, no doubts that quantitative ones will be available pretty soon.

To conclude this subchapter, DTd features principal things of thermodynamics and as a ruler is as blind as that whole science: it predicts probabilities and separates possible from impossible, but never knows for sure whether the possible happens indeed.

### 7.3. Major DTd problems to solve.

One should mention clarification of the $\omega_0(\delta_j)$ and $\pi$ values in the Le Chatelier's Response equation first. That should be done via comparison to experimental data and finding meaningful members in the $\delta$ power series and their weights, thus perhaps eliminating the need in $\omega_0(\delta_j)$ at all. The researcher ought to be ready to any surprises for "we cannot demand from Nature simplicity, nor cannot we judge what in her opinion is simple" [14]. We have confirmed earlier, that the simplest analytical force-shift relationship (27) can be proven in a simple computer experiment as well as its complicated form in map (24) has indirect analogs in experimental data.

Among the basic tasks, DTd application to open systems with chemical interactions between them and surroundings has the highest priority. This is the way to simulate complex chemical equilibria and to create appropriate software, which from several points of view will be more correct in analysis and treatment of real chemical systems than currently used classical based simulation programs. Solving this task may (and the author is sure, that it will) offer a purely thermodynamic picture of chemical oscillations in BZ, Brey-Liebhafsky and other oscillating reactions. Next very important problem is to compare the LCR approach with the results of usage explicit thermodynamic forces in such systems like electrochemical, photo-chemical and lasers as quasi-chemical systems (for some preliminary results see [35]). And, of course, direct experiments are extremely welcome.

We have touched only the surface of the DTd electrochemical application. Some more electrochemical tasks look quite feasible for successful DTd trial. Among them is discrete thermodynamic model of overpotential as manifestation of indifferent (or quasi-indifferent) equilibrium (even being caused by kinetic reasons). Another potential problem is the Batler-Volmer equation with new meaning of the charge transfer coefficients; in the DTd one can easily prove that cathode and anode shifts are related as $\delta_{jcat}=1-\delta_{jan}$, thus confirming traditional, but not theoretically justified relationship between the transfer coefficients; this is a new method to calculate them.

Further, as it was already mentioned, developed so far discrete thermodynamics describes chemical equilibria as equilibrium stationary states. Non-equilibrium stationary states represent a separate task, whose solution is also within capability of the new theory. Those equilibria may be approached by including in the basic map thermodynamic forces, pertaining to non-compensated fluxes; the extended map should allow us to forecast the states in the wide variety of systems on the move.

May the readers re-evaluate priorities and find out other problems to solve, particularly in their areas of interests.

**Appendix A.**

**Simulation Method and Software.**

Simulation software, created by the System Dynamics Research Foundation (SDRF) to solve the DTd problems, runs iterations with $\tau$ in the base. It moves the electronic image of chemical system along the loci of solutions to map (24) and prepares data to print out in $\delta$ vs. $\tau$ coordinates; the results may be recalculated to another reference frame on the user's command. The input includes such information on the system as the chemical reaction parameters - stoichiometric equation, standard change of Gibbs' free energy, pressure and temperature; the system parameters - initial amounts of participants and complexity factor $\pi$; iteration parameters - number of "external" iteration steps, t (usually 10,000), defining the iteration step as 1/t, number of "internal" search/iteration steps, $\alpha$ (usually 50), and precision $\varepsilon$ (most often 0.015) in finding zeroes of map (24). First, the software calculates $\eta$, corresponding to given $\Delta G^0$, T and initial composition, and populates the array of the map (24) logarithmic terms within the range (0<$t_i$<t+1). Then iterations start by setting next value of $t_i$ and mapping it onto the running value of $\tau(t_i)$ ("external" iteration), and then this process proceeds with the step $1/(t \cdot \alpha)$ to find the $\delta$ value and to test it for validity of map (24); it is accepted as the member of $\delta$–array if the map is zero, the code automatically provides for opposite signs of the map terms. The iterations continue until the number of steps exceeds the $\alpha \cdot t$ product. The software may run through ca 50 million steps in one iteration cycle on an average PC within a matter of minute or less, which is quite enough for all feasible tasks.